\documentclass[preprintnumbers,a4paper,twocolumn,floats,aps,nofootinbib,superscriptaddress]{revtex4}
\usepackage{graphicx}
\usepackage{amsmath,amssymb,amsfonts}

\newcommand{\Neff}{N_{\rm{eff}}} 
\newcommand{\He}{^4\rm{He}} 
\newcommand{\Li}{^7\rm{Li}} 
\newcommand{\DHy}{\text{D/H}}
\newcommand{\PArthENoPE}{\texttt{PArthENoPE}}
\newcommand{\CosmoMC}{\texttt{CosmoMC}} 
\newcommand{\CAMB}{\texttt{CAMB}} 
\newcommand{\NeffCMB}{\left. N_{\rm{eff}} \right|_{\rm{CMB}}} 
\newcommand{\NeffBBN}{\left. \Neff \right|_{\rm{BBN}}} 
\newcommand{\NeffYp}{\left. \Neff[Y_p] \right|_{\rm{BBN}}}

\begin{document}

\title{
Increasing $\Neff$ with particles in thermal equilibrium with neutrinos
}

\preprint{IPPP/12/44}
\preprint{DCPT/12/88}

\author{C\'eline B\oe hm}
 \email[]{c.m.boehm@durham.ac.uk}
\affiliation{Institute for Particle Physics Phenomenology, Durham University, South Road, Durham, DH1 3LE, United Kingdom}
\affiliation{LAPTH, U. de Savoie, CNRS,  BP 110,
  74941 Annecy-Le-Vieux, France}
\author{Matthew J.~Dolan}
\email[]{m.j.dolan@durham.ac.uk}
\author{Christopher McCabe}
\email[]{christopher.mccabe@durham.ac.uk}

\affiliation{Institute for Particle Physics Phenomenology, Durham University, South Road, Durham, DH1 3LE, United Kingdom}

\begin{abstract} 
Recent work on increasing the effective number of neutrino species ($\Neff$) in the early universe has focussed on introducing extra relativistic species (`dark radiation'). We draw attention to another possibility: a new particle of mass $\lesssim10$ MeV that remains in thermal equilibrium with neutrinos until it becomes non-relativistic increases the neutrino temperature relative to the photons. We demonstrate that this leads to a value of $\Neff$ that is greater than three and that $\Neff$ at CMB formation is larger than at BBN. We investigate the constraints on such particles from the primordial abundance of helium and deuterium created during BBN and from the CMB power spectrum measured by ACT and SPT and find that they are presently relatively unconstrained. We forecast the sensitivity of the Planck satellite to this scenario: in addition to dramatically improving constraints on the particle mass, in some regions of parameter space it can discriminate between the new particle being a real or complex scalar.  
 \end{abstract} 

\date{\today}
\maketitle

\section{Introduction}

Observations of the abundance of light nuclei produced during big-bang nucleosynthesis (BBN) and precision measurements of temperature anisotropies in the cosmic microwave background (CMB) provide unique windows into the particle content of the early universe. Both epochs are sensitive to the expansion rate of the early universe, which depends on the number of relativistic species in thermal equilibrium. The number of relativistic species is usually parameterised in terms of $\Neff$, the number of effective neutrino species with a late-time neutrino-to-photon temperature ratio \mbox{$T_{\nu}/T_{\gamma}=(4/11)^{1/3}$}. In the standard cosmological model with three neutrino species, this temperature ratio is slightly higher than $(4/11)^{1/3}$ due to partial reheating of the neutrinos when electrons and positrons annihilate, leading to $\Neff=3.046$ \cite{Mangano:2001iu, Mangano:2005cc}. A measurement of $\Neff$ greater than this would be a clear sign of new physics.

BBN imposes limits on $\Neff$ at a photon temperature around $1-0.1$ MeV, since increasing $\Neff$ leads to a higher abundance of primordial helium and deuterium~\cite{Sarkar:1995dd}. While the determination of these light nuclei from astrophysical observations is not yet completely settled, the latest data show a preference for a value of $\Neff$ greater than three. For instance, ref.~\cite{Izotov:2010ca} finds \mbox{$\Neff=3.7^{+0.8}_{-0.7}\,(2\sigma)$} from their inferred abundance of primordial $\He$.  An independent determination of $\Neff$ can be made at photon decoupling (when the photon temperature is around 1 eV) from the Silk damping \cite{Silk:1967kq} tail of the CMB power spectrum. The Atacama Cosmology Telescope (ACT)~\cite{Dunkley:2010ge} and the South Pole Telescope (SPT)~\cite{Keisler:2011aw} have measured this damping tail and, in combination with data from the Wilkinson Microwave Anisotropy Probe (WMAP)~\cite{Komatsu:2010fb}, baryon acoustic oscillations (BAO)~\cite{Percival:2009xn} and the Hubble constant~($H_0$)~\cite{Riess:2009pu, Riess:2011yx}, also find a preference for a value of $\Neff$ greater than three, obtaining \mbox{$\Neff=4.6\pm0.8\,(1\sigma)$} and \mbox{$\Neff=3.9\pm0.4\,(1\sigma)$} respectively. Furthermore, various independent analyses of recent cosmological data find evidence for $\Neff>3$ at \mbox{95\%~CL~\cite{ Smith:2011es, Hamann:2011ge, Archidiacono:2011gq, Hamann:2011hu, Nollett:2011aa}.}

Given these hints for extra energy density in the early universe and bearing in mind the expectation that the Planck experiment will soon provide a significantly improved measurement of $\Neff$ at photon decoupling (e.g.~see~\cite{Galli:2010it}), we consider it a pertinent time to consider models in which $\Neff$ is increased. The conventional way is to introduce extra `dark' radiation, which leads to the same value of $\Neff$ at BBN and photon decoupling (see~e.g.~\cite{Nakayama:2010vs,Feng:2011uf,Dreiner:2011fp}).  While it is usually assumed that $\Neff$ does not change between BBN and photon decoupling, intriguingly, the experimental data are consistent with a slightly larger value of $\Neff$ at photon decoupling. This increase in $\Neff$ after BBN can be achieved through decays to dark radiation during or after BBN \mbox{(see~e.g.~\cite{Ichikawa:2007jv, Fischler:2010xz, Hasenkamp:2011em,  Menestrina:2011mz,Hooper:2011aj, Bjaelde:2012wi})} or from primordial gravitational waves~\cite{Smith:2006nka}.

In this paper we explore another scenario that has received less attention in the literature, which predicts a value of $\Neff$ greater than three, and moreover, predicts that $\Neff$ at photon decoupling is larger than the value during BBN. As we will show in section~\ref{Section:Current}, $\Neff$ scales as $(T_{\nu}/T_{\gamma})^4$ so increasing the neutrino-to-photon temperature ratio leads to an increase in $\Neff$. A relative increase in $T_{\nu}/T_{\gamma}$ can be achieved by reducing the photon temperature relative to the neutrino temperature or by increasing the neutrino temperature relative to the photon temperature. A decrease in $T_{\gamma}$ can be obtained through the production of hidden photons \cite{Jaeckel:2008fi,Foot:2011ve} while a new light mediator may lead to a change in either $T_{\gamma}$ or~$T_{\nu}$~\cite{Blennow:2012de}. In this paper we follow the example of the standard cosmological model to increase $T_{\nu}$: in the standard cosmological model, the photons are hotter than the neutrinos because the transfer of entropy from the electrons and positrons to the photons (when they become non-relativistic) happens after the neutrinos decouple from the electromagnetic plasma at $T_{\rm{D}}\approx2.3$~MeV~\cite{Enqvist:1991gx}. In an analogous way, we show that an additional `generic' particle $\chi$ that remains in thermal equilibrium solely with the neutrinos after they decouple from the electromagnetic plasma and until it is non-relativistic, reheats the neutrinos relative to the electromagnetic plasma and as a result, leads to a higher value for $\Neff$. Requiring that the neutrino reheating, which happens when $\chi$ becomes non-relativistic and transfers its entropy to the neutrinos, happens after neutrino decoupling implies that $\chi$ must have a mass $m_{\chi}\lesssim {\rm{few}}\cdot T_{\rm{D}}\sim10$ MeV. 

The goal of this paper is to investigate the current and future constraints on $\chi$ from the inferred values of $\Neff$ from BBN and the CMB . The only condition that we impose on the `generic' particle $\chi$ is that the neutrino-$\chi$ interaction rate is sufficiently high that they remain in thermal equilibrium until $\chi$ is non-relativistic. Therefore, the constraints we derive may apply to, for instance, light dark matter particles that obtain their abundance from thermal freeze out (see e.g.~\cite{Boehm:2000gq, Boehm:2002yz, Boehm:2006mi,Farzan:2009ji}) or light mediators that have large couplings to neutrinos (see e.g.~\cite{Beacom:2004yd, Hannestad:2004qu}). The parameter that determines $\Neff$ is $m_{\chi}$ as this dictates the additional energy density, so it is this parameter that we will constrain. During BBN, the photon temperature is similar to $m_{\chi}$ so $\chi$ makes a direct contribution to $\Neff$. In addition, there may also be an indirect contribution from the increase in $T_{\nu}/T_{\gamma}$. At photon decoupling, we assume $\chi$ is non-relativistic so its direct contribution to $\Neff$ is Boltzmann suppressed; the increase in $\Neff$ arises solely from the increase in $T_{\nu}/T_{\gamma}$. As we show in section~\ref{Section:Current}, this difference in the origin of the extra energy density leads to a larger value of $\Neff$ at photon decoupling.

This paper is organised as follows. In section~\ref{Section:Current} we discuss the impact of light particles in thermal equilibrium with neutrinos on the energy density and neutrino-to-photon temperature ratio (some results are fully derived in an Appendix). Using these results, we first find the effect of $\chi$ on the abundance of primordial nuclei produced during BBN. This has previously been studied in~\cite{Kolb:1986nf, Serpico:2004nm}. Here, we independently calculate the primordial abundance of $\He$ and $\DHy$ as a function of $m_{\chi}$ and compare with recent experimental measurements. Our calculation uses a more recent BBN code than \cite{Kolb:1986nf, Serpico:2004nm} and includes the most recent values of the neutron lifetime and baryon density. We then calculate the value of $\Neff$ as a function of $m_{\chi}$ at photon decoupling and compare with the experimental values inferred by ACT and SPT. Turning to consider experimental results from the near future, in section~\ref{Section:Planck} we forecast the constraints that will soon be placed on $\chi$ from Planck's measurement of $\Neff$. Finally, we conclude in section~\ref{Section:Conclusion}.

\section{Current constraints} \label{Section:Current}

The total energy density of the universe $\rho_{\rm{R}}$ is usually parameterised in terms of the energy density of photons $\rho_{\gamma}$, and the  effective number of neutrinos $\Neff$ with the usual neutrino-to-photon temperature ratio $T^0_{\nu}/T_{\gamma}$ (we define this ratio more carefully below)
\begin{equation}
\label{eq:phoR}
\rho_{\rm{R}}=\rho_{\gamma}\left[1+\frac{7}{8} \left(\frac{T^0_{\nu}}{T_{\gamma}}\right)^4 \Neff \right]\;.
\end{equation}

We are interested in the case where an additional particle $\chi$ with $g_{\chi}$ internal degrees of freedom and mass $m_{\chi}$ is in thermal equilibrium with the neutrinos. As the particles are in thermal equilibrium, they share a common temperature $T_{\nu}$, which will in general be different from $T_{\nu}^0$. The resulting energy density of $N_{\nu}$ neutrinos and the additional particle $\chi$ is 
\begin{equation}
\label{eq:rhonuchi}
\rho_{\nu:\chi}=\rho_{\gamma}\cdot \frac{7}{8}\,\left(\frac{T_{\nu}}{T_{\gamma}}\right)^4\,\left[N_{\nu}+\frac{g_{\chi}}{2}I(y_{\nu}) \right]\, ,
\end{equation}
where for convenience, we have defined $y_{\nu}\equiv m_{\chi}/T_{\nu}$ and the function
\begin{align}
\label{eq:Iy}
I(y)=\frac{120}{7 \pi^4} \int^{\infty}_y d\xi \frac{\xi^2 \sqrt{\xi^2-y^2}}{e^{\xi}\pm1}\;,
\end{align}
which takes the limits $I(y\rightarrow\infty)= 0$ and \mbox{$I(y\rightarrow0)= 1 \,(8/7)$} for fermions (bosons) (as usual, in eq.~\eqref{eq:Iy}, $-1$ pertains to bosons and $+1$ to fermions). Comparing eq.~\eqref{eq:rhonuchi} with the definition of $\Neff$ in eq.~\eqref{eq:phoR}, we see that
\begin{equation}
N_{\rm{eff}}(y_{\nu})=\left(\frac{T_{\nu}^0}{T_{\gamma}}\right)^{-4}\left(\frac{T_{\nu}}{T_{\gamma}}\right)^{4} \left[N_{\nu}+\frac{g_{\chi}}{2} I(y_{\nu})  \right]\;.
\label{eq:Neff1}
\end{equation}

In the Appendix, we use the conservation of entropy per-comoving-volume for particles in thermal equilibrium to show that (for $T_{\gamma}\leq T_{\rm{D}}$)
\begin{equation}
\frac{T_{\nu}}{T_{\gamma}}=\left(\left. \frac{g_{\star s: \nu}}{g_{\star s: \gamma}}\right|_{T_{\rm{D}}} \frac{g_{\star s: \gamma}}{g_{\star s: \nu}} \right)^{1/3}\;,
\label{eq:Tratio}
\end{equation}
where $g_{\star s: \nu}$ and $g_{\star s: \gamma}$  are the effective number of relativistic degrees of freedom in the neutrino and electromagnetic sectors respectively and $\left.\right|_{T_{\rm{D}}}$ indicates that $g_{\star s}$ should be evaluated at the neutrino decoupling temperature $T_{\rm{D}}$. In the absence of $\chi$, $g_{\star s: \nu}$ remains constant after decoupling, hence the usual neutrino-to-photon ratio is
\begin{equation}
\frac{T_{\nu}^0}{T_{\gamma}}=\left(\frac{g_{\star s: \gamma}}{\left.g_{\star s: \gamma} \right|_{T_{\rm{D}}}}\right)^{1/3}\;.
\end{equation}
More generally, we see that 
\begin{equation}
\label{eq.TnuTnu0}
\left(\frac{T_{\nu}}{T_{\gamma}}\right)=\left(\frac{T^0_{\nu}}{T_{\gamma}}\right) \left(\frac{\left.g_{\star s: \nu} \right|_{T_{\rm{D}}}}{g_{\star s: \nu}}\right)^{1/3}\;.
\end{equation}
If $\chi$ becomes non-relativistic after decoupling, \mbox{$g_{\star s: \nu}<\left.g_{\star s: \nu} \right|_{T_{\rm{D}}}$} and as a result, the neutrino-to-photon temperature ratio increases above its usual value. From eq.~\eqref{eq:Neff1}, we see that this leads to an increase in $\Neff$.

\begin{figure}[t]
\begin{center}
{
 \includegraphics[width=0.99\columnwidth]{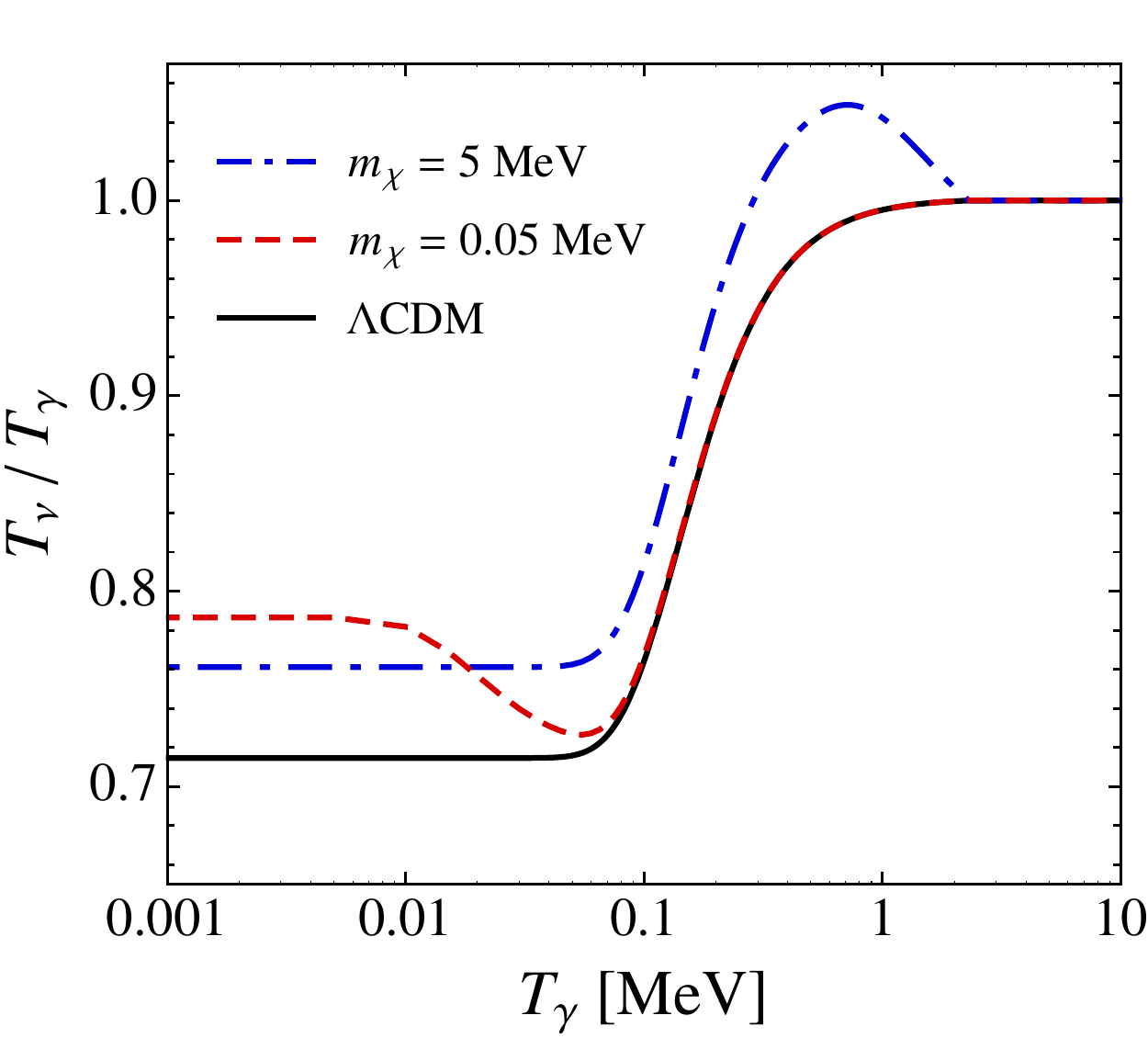}
}
\end{center}
\caption{The evolution of $T_{\nu}/T_{\gamma}$ with $T_{\gamma}$ in the standard concordance model `$\Lambda$CDM' (black solid) and when there is an additional Majorana fermion in thermal equilibrium with neutrinos with mass \mbox{$m_{\chi}=5$ MeV} (blue dot-dashed) and \mbox{$m_{\chi}=0.05$~MeV} (red dashed). Neutrino reheating occurs when $T_{\gamma}\sim m_{\chi}$.
}
\label{fig:TnuTg}
\end{figure}

Finally, in the Appendix we show that eq.~\eqref{eq.TnuTnu0} can be written as
\begin{equation}
\label{eq:TnuTnu0F}
\left(\frac{T_{\nu}}{T_{\gamma}}\right)=\left(\frac{T^0_{\nu}}{T_{\gamma}}\right)\left[ \frac{\left.N_{\nu}+\frac{g_{\chi}}{2}F\left(y_{\nu}\right|_{T_{\rm{D}}}\right)}{N_{\nu}+\frac{g_{\chi}}{2} F(y_{\nu})}    \right]^{1/3}\;,
\end{equation}
where $\left.y_{\nu}\right|_{T_{\rm{D}}}=m_{\chi}/T_{\rm{D}}$ and we have defined the function
\begin{align}
\label{eq:F}
F(y)=\frac{30}{7 \pi^4}\int^{\infty}_y d\xi \frac{(4 \xi^2-y^2)\sqrt{\xi^2-y^2}}{e^{\xi}\pm1}\;,
\end{align}
which takes the limits $F(y\rightarrow\infty)= 0$ and \mbox{$F(y\rightarrow0)= 1 \,(8/7)$} for fermions (bosons). It is this form of $T_{\nu}/T_{\gamma}$ that we use in our calculations. In figure~\ref{fig:TnuTg} we show how $T_{\nu}/T_{\gamma}$ evolves with $T_{\gamma}$ for three cases: the black solid line shows the result from the standard concordance model with three neutrino species, in which the photon reheating occurs when $T_{\gamma}\sim m_{e}$; the blue dot-dashed and red dashed lines show the evolution when an additional Majorana fermion with mass $m_{\chi}=5$~MeV and $m_{\chi}=0.05$~MeV is in thermal equilibrium with the neutrinos. We see that the neutrino reheating occurs when $T_{\gamma}\sim m_{\chi}$. For the $m_{\chi}=5$~MeV case, some of the reheating occurs for $T_{\gamma}\geq T_{\rm{D}}$ during which the photons and neutrinos are reheated equally. In comparison, for the $m_{\chi}=0.05$~MeV case, all of the reheating occurs for $T_{\gamma}\leq T_{\rm{D}}$ so only the neutrinos are reheated and therefore, the ratio of $T_{\nu}/T_{\gamma}$ is larger than the $m_{\chi}=5$~MeV case at late times (small $T_{\gamma}$).

In the following subsections, we use the above formulae to explore the implications of $\chi$ on BBN and the CMB.

\subsection{Effect on BBN} \label{Section:BBN}

The abundance of $\He$, expressed through the mass fraction $Y_p$, has long been recognised as a probe of the energy density present during BBN \cite{Wagoner:1966pv}. In the early universe the ratio of the neutron-to-proton number densities ($n_n/n_p$) is kept in thermal equilibrium through the weak interactions until the expansion rate becomes comparable to the weak interaction rate at $T\sim0.7$ MeV.  Increasing the expansion rate, by for instance introducing additional relativistic radiation, leads to a larger value of $n_n/n_p$ at freeze out. Since essentially all of the neutrons present at $T\sim0.1$ MeV are synthesised into $\He$, introducing additional relativistic radiation therefore leads to a larger abundance of $\He$. Similarly, increasing the expansion rate means that the reactions depleting the abundance of $\rm{D}$ freeze out earlier, leading to an increased abundance. The reactions that deplete $\rm{D}$ are more sensitive to the baryon-to-photon number density $\eta$ and have historically been used to infer its value. However,
if we use the value of $\eta$ inferred by WMAP, then the number ratio $\DHy$ can be used along with $Y_p$ as a measure of the energy density present during BBN, a point emphasised in \cite{Nollett:2011aa}.  Finally, the observed abundance of $\Li$ remains $\sim 5 \sigma$ away from the theoretical prediction, the so-called `lithium problem' (for a recent review, see \cite{Fields:2012jf}). The modifications that we propose do not significantly alter the $\Li$ abundance, therefore, in this work we directly compare our theoretical predictions with the inferred values $Y_p$ and $\DHy$ from astrophysical measurements and make only passing reference to $\Li$. 

The impact on the primordial abundances of light nuclei from a generic particle $\chi$ that maintains thermal equilibrium with neutrinos throughout BBN has previously been considered in \cite{Kolb:1986nf, Serpico:2004nm}. Here we independently calculate $Y_p$ and $\DHy$ using a modified version of the $\PArthENoPE$ BBN code~\cite{Pisanti:2007hk}. We use the latest measurements of the baryon density $\Omega_{\rm{b}} h^2=0.0223$, deduced from a joint analysis of SPT+WMAP7+$H_0$+BAO~\cite{Keisler:2011aw}, and the PDG value of the neutron lifetime \mbox{$\tau_n=880.1$~s}~\cite{Beringer2012}.

In the spirit of \cite{Serpico:2004nm}, we introduce a small temperature dependent parameter
\begin{align}
\delta(T)&\equiv1-\frac{T_{\nu}^0(T)}{T_{\nu}(T)}\;,
\end{align}
and perturb the $\PArthENoPE$ code in order to take into account the effects from the modified neutrino-to-photon temperature ratio. We use eq.~\eqref{eq:TnuTnu0F} to calculate $T_{\nu}^0/T_{\nu}$. Typically, \mbox{$\delta\sim0.01$} and is always smaller than $\sim0.1$. A non-zero value of $\delta$ increases the energy density of the neutrinos and enters into the phase-space of the weak interaction rates that determine $n_n/n_p$. The $\PArthENoPE$ interaction rates $\tilde{\Gamma}_{n \rightarrow p}$ and $\tilde{\Gamma}_{p \rightarrow n}$ include finite mass, QED radiative and finite temperature corrections. We calculate corrections to these rates in the Born approximation as an expansion in $\delta$ such that the total rate for $n\rightarrow p$ is given by
\begin{equation}
\Gamma_{n\rightarrow p}^{\rm{total}}=\tilde{\Gamma}_{n \rightarrow p}+\epsilon_{n1}(T)\delta +\epsilon_{n2}(T)\delta^2+\epsilon_{n3}(T)\delta^3\;,
\end{equation}
where $\epsilon_{ni}(T)$ fits the change in the rates to an accuracy better than a few percent. A similar expression holds for the $p \rightarrow n$ rate. Finally, we also include the extra contribution to the energy density from $\chi$, expressed in eq.~\eqref{eq:rhonuchi}.

\begin{figure}[t]
\begin{center}
{
 \includegraphics[width=0.99\columnwidth]{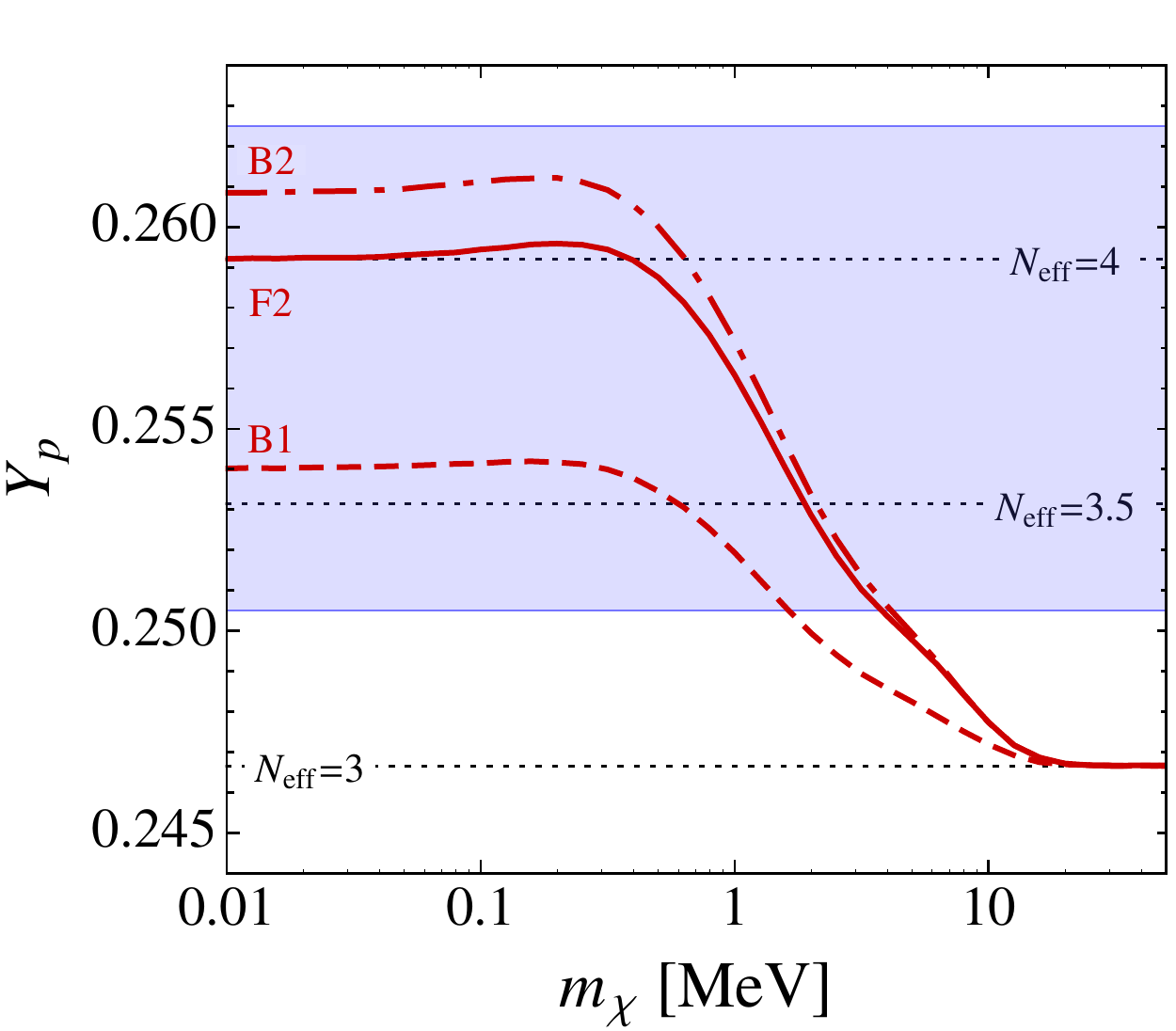}
  \includegraphics[width=0.953\columnwidth]{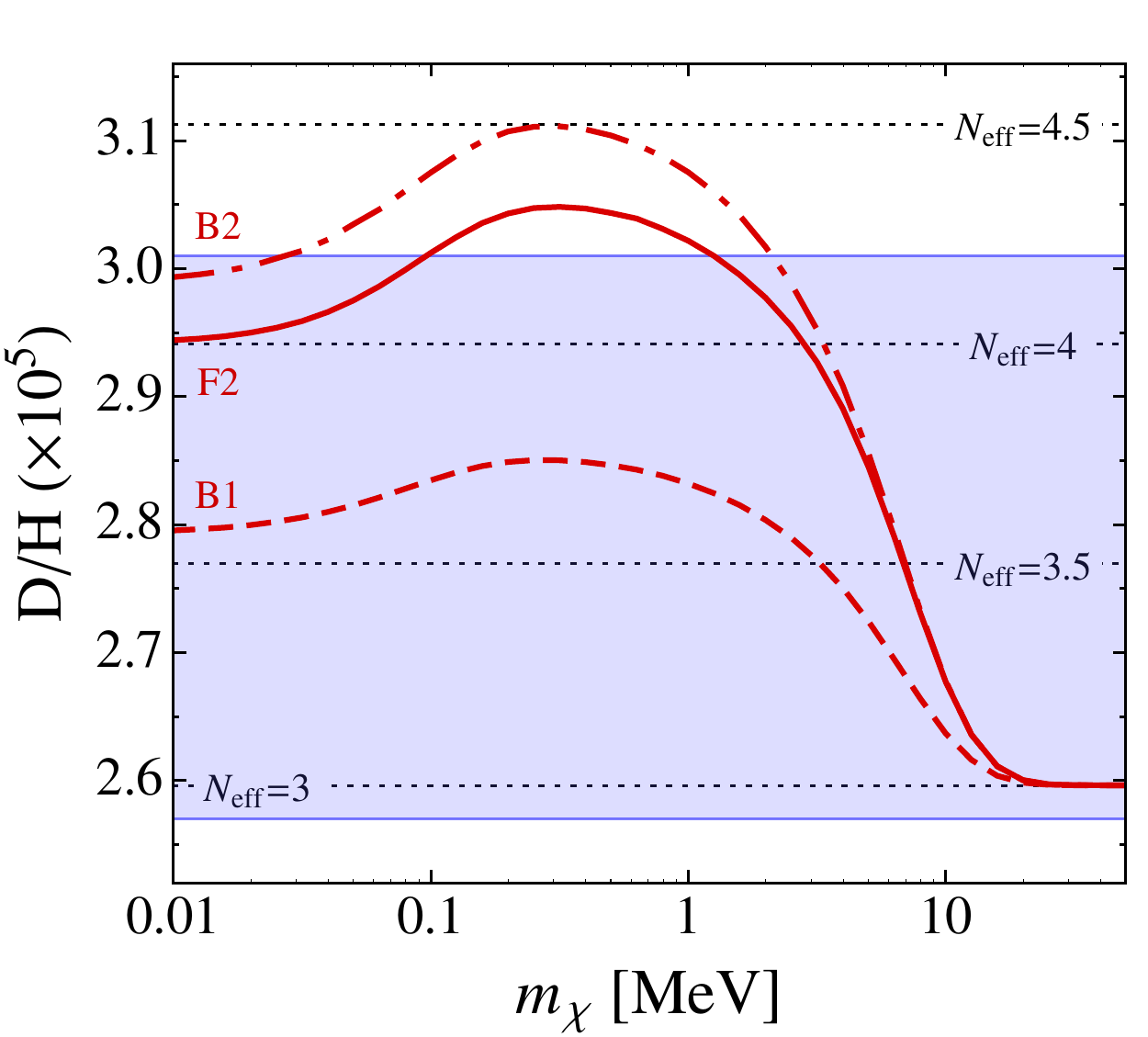} \hspace{-6mm}
}
\end{center}
\caption{The red dot-dashed, solid and dashed lines show the predictions for $Y_p$ (upper panel) and $\DHy$ (lower panel) for a complex scalar (B2), Majorana fermion (F2) and real scalar (B1) respectively. The blue shaded region indicates the $1\sigma$ region for $Y_p$ from \cite{Izotov:2010ca} (with statistical and systematic errors combined linearly) and the $1\sigma$ weighted mean of \mbox{$\DHy$} from \cite{Fumagalli:2011iw}. The black dotted lines show the values of $Y_p$ and $\DHy$ for the indicated values of $\Neff$.
}
\label{fig:YpDH}
\end{figure}

The red dot-dashed, solid and dashed lines in figure~\ref{fig:YpDH} show our results for a complex scalar (B2), Majorana fermion (F2) and real scalar (B1) respectively as a function of $m_{\chi}$. The upper and lower panels show the values of $Y_p$ and $\DHy$ respectively and are in good qualitative agreement with those found in \cite{Kolb:1986nf, Serpico:2004nm}, with slight differences due to the updated parameter values that we use. Although not shown here, our prediction for $\Li$ is similar to that in \cite{Serpico:2004nm}. For reference, the black dotted lines show the predicted values of $Y_p$ and $\DHy$ for the indicated values of $\Neff$. As we would expect, for $m_{\chi}\gtrsim15$ MeV, we recover the result from standard BBN: this is because $\chi$ is non-relativistic during BBN (so its contribution to the energy density is Boltzmann suppressed) and it has transferred its entropy to the neutrinos before they decouple from the photons, meaning that the standard neutrino-to-photon temperature ratio relation is maintained ($g_{\star s: \nu}=\left.g_{\star s: \nu} \right|_{T_{\rm{D}}}$ in eq.~\eqref{eq.TnuTnu0}). For \mbox{$m_{\chi}\lesssim0.05$ MeV}, we asymptote to the result expected from a massless particle as $\chi$ remains relativistic throughout all of BBN. The values of $Y_p$ and $\DHy$ for intermediate values of $m_{\chi}$ are a result of the direct contribution to the energy density from $\chi$ and the contribution from the modified neutrino-to-photon temperature ratio.

Recent inferences of $Y_p$ from observations of metal-poor H II regions have been slightly higher than results from the past decade. For instance, while refs.~\cite{Olive:2004kq} and~\cite{Peimbert:2007vm} found \mbox{$Y_p=0.249\pm0.009$} and \mbox{$Y_p=0.2477\pm0.0029$} respectively, more recently, refs.~\cite{Izotov:2010ca}, \cite{Aver:2010wq}, \cite{Aver:2010wd} and~\cite{Aver:2011bw} found $Y_p=0.2565\pm0.0010(\rm{stat.})\pm0.0050(\rm{syst.})$, $Y_p=0.2561\pm0.0108$, $Y_p=0.2573^{+0.0033}_{-0.0088}$ and $Y_p=0.2534\pm0.0083$ respectively, in good agreement with each other. The agreement of these recent results is perhaps not surprising as they originate from independent analyses of subsets of the data compiled in ref.~\cite{Izotov:2007ed}. However, all of these numbers serve to highlight that the value of $Y_p$ is dominated by systematic errors. Therefore, following \cite{Steigman:2012ve}, we take the value of $Y_p$ from ref.~\cite{Izotov:2010ca} with the statistical and systematic errors combined linearly: $Y_p=0.2565\pm0.006$. The blue shaded region in the upper panel of figure~\ref{fig:YpDH} shows this $1\sigma$ region.

The experimental value for the number ratio $\DHy$ shows a large spread with the dispersion in the measurements well in excess of the quoted errors. For example, measurements range from $\DHy\times10^5=1.17^{+0.48}_{-0.34}$ in~\cite{2010MNRAS.405.1888S} to $\DHy\times10^5=3.98^{+0.70}_{-0.59}$ in~\cite{Burles:1997fa}. Here we take as a conservative value \mbox{$\DHy\times10^5=2.78^{+0.23}_{-0.21}$}, the weighted mean calculated from eight independent samples in \cite{Fumagalli:2011iw}. The blue shaded region in the lower panel of figure~\ref{fig:YpDH} shows this $1\sigma$ region. It is clear from figure~\ref{fig:YpDH} that the error on the measurements of $Y_p$ and $\DHy$ are sufficiently large that all values of $m_{\chi}$ are currently consistent with the data at just over $1\sigma$. Although the measured errors of $Y_p$ are unlikely to improve considerably in the near future, recently, a precise measurement of $\DHy$ was presented with an error $\sim4$ times smaller than the previous best \cite{Pettini:2012ph}. With this measurement, they find that the weighted mean from ten independent samples is \mbox{$\DHy\times10^5=2.63\pm0.12$} (where the shift in the weighted mean and reduction of the error is largely due to the new measurement in \cite{Pettini:2012ph}). Comparing with the lower panel of figure~\ref{fig:YpDH}, we see that this result disfavours $m_{\chi}\lesssim4$ MeV at $2\sigma$ for a Majorana fermion and complex scalar while still allowing all values of $m_{\chi}$ for a real scalar. With further observations from metal-poor damped Lyman alpha systems of the type considered in~\cite{Pettini:2012ph}, the large dispersion in the measurements of $\DHy$ would decrease and provide a strong, reliable constraint on this scenario.

\subsection{Effect on the CMB} \label{section:CMB}

\begin{figure}[t]
\begin{center}
{
\includegraphics[width=0.99\columnwidth]{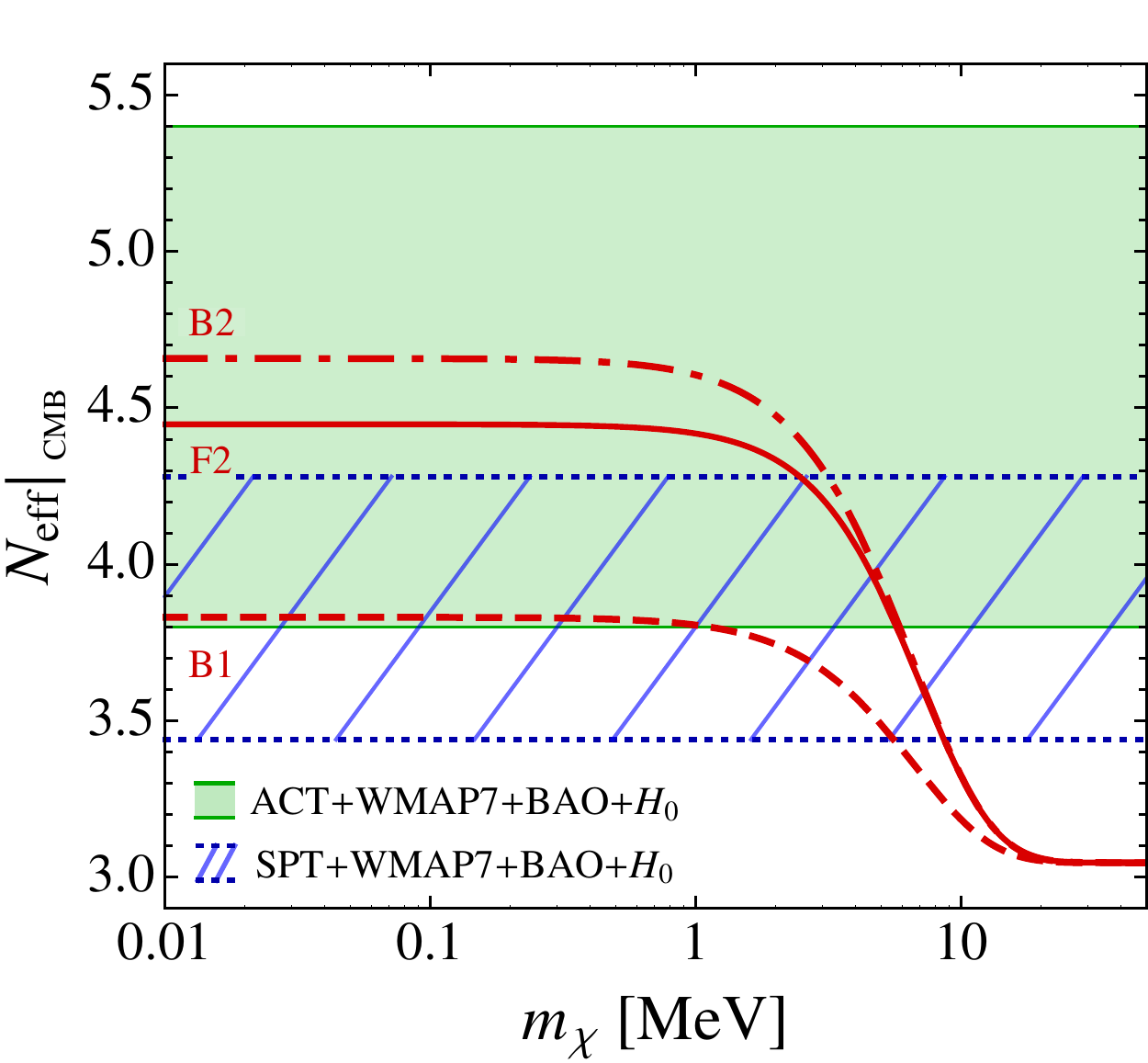}
}
\end{center}
\caption{The red dot-dashed, solid and dashed lines show $N_{\rm{eff}}$ at photon decoupling ($\NeffCMB$) as a function of $m_{\chi}$ for a complex scalar (B2), Majorana fermion (F2) and real scalar respectively. The shaded green and blue hatched regions show the $1\sigma$ range of $\Neff$ determined from ACT and SPT (in combination with data from WMAP7, BAO and $H_0$) respectively.}
\label{fig:NeffCMB}
\end{figure}

An independent determination of $\Neff$ can be made from the CMB, which is at a much lower photon temperature than found during BBN. For the values of $m_{\chi}$ that we consider, $\chi$'s direct contribution to the energy density is Boltzmann suppressed at photon decoupling \mbox{($T_{\rm{\gamma\text{-}dec}}\sim1$ eV)}: that is, $I(m_{\chi}/T_{\rm{\gamma\text{-}dec}})\simeq 0$ in eq.~\eqref{eq:rhonuchi}. However, assuming that $\chi$ remains in thermal equilibrium with neutrinos until it is non-relativistic, we can calculate its indirect contribution from the increase in the neutrino-to-photon temperature ratio from eq.~\eqref{eq:TnuTnu0F} and the resulting value of $\Neff$ from eq.~\eqref{eq:Neff1}. At photon decoupling we have $F(m_{\chi}/T_{\rm{\gamma\text{-}dec}})\simeq0$ so $\Neff$ is given~by
\begin{equation}\label{NeffCMB1}
\left. \Neff \right|_{\rm{CMB}} \simeq 3.046 \left[ 1+ \frac{g_{\chi}}{2}\frac{\left. F(y_{\nu}\right|_{T_{\rm{D}}})}{3.046}\right]^{4/3}\;,
\end{equation}
where we have set $N_{\nu}=3.046$ \cite{Mangano:2001iu}. In figure~\ref{fig:NeffCMB} we show $N_{\rm{eff}}$ as a function of $m_{\chi}$ for a complex scalar (B2), Majorana fermion (F2) and real scalar (B1) in red dot-dashed, solid and dashed lines respectively. For $m_{\chi}\gtrsim15$ MeV we recover the usual value for $\Neff$ because the reheating effect from $\chi$ takes place before $T_{\rm{D}}$, so the neutrinos and photons are both reheated equally. For \mbox{$m_{\chi}\lesssim2$ MeV}, $\Neff$ remains constant because the neutrinos are maximally reheated for all of these values of $m_{\chi}$. This is because $\chi$ becomes non-relativistic after $T_{\rm{D}}$ and is fully non-relativistic by photon decoupling, therefore, all entropy from $\chi$ is transferred to the neutrinos. Naively, we might have expected $\Neff$ to asymptote to four when $\chi$ is a light Majorana fermion. That this does not happen can be traced to the fact that the term in square brackets in eq.~\eqref{NeffCMB1} is raised to a power $4/3$, rather than $1$. This in turn follows from entropy conservation; the factor of 3 follows from the fact that $S\sim g_{\star s} T^3$ is conserved, rather than $g_{\star s} T^4$.

Experimentally, $\left. \Neff \right|_{\rm{CMB}}$ is determined from the CMB power spectrum at high multipole $\ell$ (`the Silk damping tail'). Increasing $\Neff$ increases the energy density, which in turn increases the expansion rate. As discussed in~\cite{Hou:2011ec}, an increased expansion rate leads to increased Silk damping which reduces the power in damping tail of the CMB power spectra.  Currently, ACT~\cite{Dunkley:2010ge} and SPT~\cite{Keisler:2011aw} provide the most precise measurements of the CMB power spectra at high $\ell$ (up to $\ell\sim3000$) and are thus able to set the best limits on $\NeffCMB$ from the CMB power spectrum. In figure~\ref{fig:NeffCMB}, we show the current $1\sigma$ constraints on $\NeffCMB$ from ACT (shaded green region) and SPT (blue hatched region) when their respective data sets are combined with data from WMAP7+BAO+$H_0$.\footnote{Strictly speaking, these constraints apply when $Y_p$ is fixed to the standard BBN value with $\Neff$ massless fermions. However, from table~\ref{tab:forecast}, we see that the constraints on $\NeffCMB$ (from SPT+WMAP7+BAO+$H_0$) do not significantly change when $Y_p$ is left as a free parameter.} While we see that the ACT and SPT data show a preference for $\Neff>3$ and therefore, $m_{\chi}\lesssim 10$ MeV, all values of $m_{\chi}$ are currently consistent with the data at $2 \sigma$. As a result, no strong conclusions on the preferred values of $m_{\chi}$ can currently be drawn from the data.

\subsection{Comparison}\label{Se:comparison}

\begin{figure}[t]
\begin{center}
{
\includegraphics[width=0.99\columnwidth]{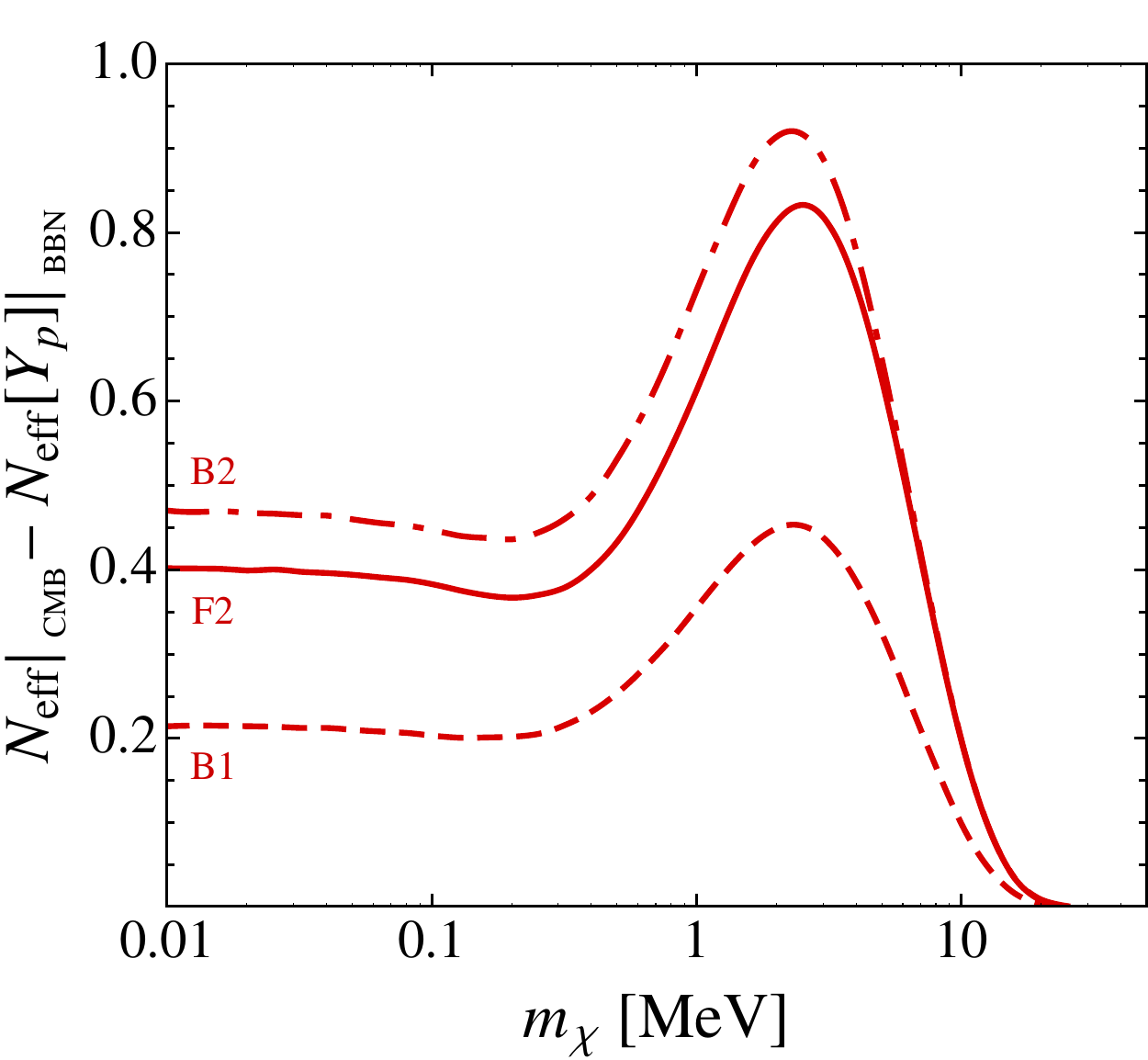}
}
\end{center}
\caption{The difference in $N_{\rm{eff}}$ at photon decoupling and BBN (inferred from the value of $Y_p$).  The red dot-dashed, solid and dashed lines show the prediction for a complex scalar (B2), Majorana fermion (F2) and real scalar (B1) respectively.}
\label{fig:Neffcomp}
\end{figure}

Comparing figures~\ref{fig:YpDH} and \ref{fig:NeffCMB}, we see that $\NeffBBN$ and $\NeffCMB$ differ for the same value of $m_{\chi}$. In figure~\ref{fig:Neffcomp} we explicitly show the difference in $\Neff$ between the BBN and the CMB epochs as a function of $m_{\chi}$ for a complex scalar (B2), Majorana fermion (F2) and real scalar (B1) (in red dot-dashed, solid and dashed lines respectively). Here, we follow the usual convention and use the inferred value of $\NeffBBN$ from $Y_p$ ($\NeffYp$) but note that a similar figure could be drawn in which $\NeffBBN$ is inferred from the values of $\DHy$. When a particle $\chi$ remains in thermal equilibrium with neutrinos until non-relativistic, we see that $\NeffCMB \geq \NeffYp$. The equality holds for $m_{\chi}\gtrsim15$ MeV because $\Neff$ remains at its usual value. For $m_{\chi}\lesssim0.2$ MeV, the difference is essentially constant because $\Neff$ takes the maximum value at BBN and photon decoupling. The difference in $\Neff$ is greatest for intermediate values of $m_{\chi}$. 

Finally, for completeness we mention a way to achieve $\NeffCMB \leq \NeffYp$. If $\chi$ decouples from the neutrinos while still relativistic but after BBN (implying $m_{\chi}\lesssim0.1$ MeV) we would have $\NeffYp=4$ and no reheating of the neutrinos would occur. Furthermore, if $m_{\chi}\gtrsim1$ eV, $\chi$ would not contribute to the energy density at photon decoupling since it would be non-relativistic. Therefore, $\NeffCMB=3.046$ as usual. We will not explore this possibility any further.

\section{Planck forecast}\label{Section:Planck}

In the previous section, we demonstrated in figures~\ref{fig:YpDH} and~\ref{fig:NeffCMB} that all values of $m_{\chi}$ are consistent (within $2\sigma$) with the current measurements of $\Neff$ from BBN and the CMB for a particle $\chi$, which remains in thermal equilibrium with the neutrinos after they decouple from the electromagnetic plasma and until it is non-relativistic.  However, we would ideally like to place strong constraints on $m_{\chi}$ or find evidence for such a scenario. As Planck will soon provide significantly improved measurements of the CMB temperature anisotropy, we now turn to forecast its sensitivity to our scenario. 

In section~\ref{section:CMB} we stated that $\Neff$ is determined from measurements of the damping tail of the CMB power spectrum. Due to the accuracy of Planck, we must be careful to take into account other parameters that also lead to additional damping of power on small scales as they will be degenerate with $\Neff$~\cite{Bashinsky:2003tk}. For instance, a negative running of the spectral index $n_s$ with scale $k$ is one such parameter. However, the value required to provide sufficient damping, $d n_s/d\ln k\sim-0.02$, is much larger than expected from slow-roll inflationary models and would require a rethink of inflation and the generation of perturbations \cite{Kosowsky:1995aa, Baumann:2009ds} so we will not consider it further. Other examples of such parameters like an alternative model of dark energy, are considered in~\cite{Joudaki:2012fx,Archidiacono:2012gv}. Here, we concentrate on a conventional, flat, $\Lambda$CDM model that we extend to include a parameter that has already been central to our discussion, namely, $Y_p$. Since $\He$ recombines earlier than hydrogen, increasing the $\He$ abundance (at fixed baryon density) leads to fewer free electrons during hydrogen recombination. This in turn leads to a larger photon diffusion length and as a result, less power in the CMB damping tail \cite{Steigman:2010pa}. Increasing $\Neff$ can be compensated by decreasing $Y_p$ so these parameters are degenerate~\cite{Bashinsky:2003tk}.

\begin{table}[t!]
\centering
\begin{tabular}{l c c c c}
\hline
& \raisebox{-0.7ex}{SPT+BAO+} & \multicolumn{2}{c}{\raisebox{-0.5ex}{Fiducial values}}   & \\ 
\raisebox{1.5ex}{Parameter} & \raisebox{0.7ex}{WMAP7+$H_0$} \; & \raisebox{0.5ex}{$\Lambda$CDM}&\raisebox{0.5ex}{$\Lambda$CDM$+\chi$}  & \raisebox{1.5ex}{Prior range}\\[0.3ex]  \hline \\[-1.0em]
 $100\,\Omega_{\rm{b}} h^2$ & $2.27 \pm 0.044$& 2.23 & 2.27  & $0.5\rightarrow10$ \\
 $\Omega_{\rm{DM}} h^2$ & $0.130 \pm 0.0119$ & 0.111 & 0.134 & $0.01\rightarrow0.99$ \\
 $h$& $0.73 \pm 0.022$ &  0.71 & 0.76 & $0.4 \rightarrow1.0$ \\
 $100\, \theta_S$& $1.040 \pm 0.0030$& 1.043  & 1.039 & $0.5\rightarrow10$ \\
  $\tau$ & $0.090 \pm 0.015$ & 0.09 & 0.09 & $0.01\rightarrow0.8$ \\
 $\ln[10^{10} A_{\rm{S}}]$ & $3.17 \pm 0.04$& 3.17 & 3.16 & $2.7\rightarrow4$ \\
 $n_{\rm{S}}$ & $0.989 \pm 0.013$& 0.98 & 0.99 & $0.5\rightarrow1.5$ \\
 $ f_{\nu}$ & $<0.054$& 0.008 & 0.020 & $0\rightarrow1$ \\
 $\Neff$ & $3.98 \pm 0.61$&  3.046 & 4.418 & $2\rightarrow7$\\[0.5ex]
$Y_{p}$ &\raisebox{1.5ex}{$0.2629 \pm 0.039$} &\raisebox{1.5ex}{---} & \raisebox{1.5ex}{---}& \raisebox{1.5ex}{$0\rightarrow1$} \\[-1.5ex]
  & \raisebox{0.5ex}{---} &\raisebox{0.5ex}{$0.247$}  &\raisebox{0.5ex}{$0.257$}  &\raisebox{0.5ex}{$0.22\rightarrow0.284$}   \\[0.5ex]
\hline \\[-1.0em]
$\sigma_8$ & $0.801 \pm 0.054$ & 0.803 & 0.816 & \multicolumn{1}{c}{---} \\[0.5ex]
\hline
\end{tabular}
\caption{The first column lists the cosmological parameters we sample over in our Planck forecast. The second column shows the values we obtain for these parameters when fitting to the SPT+WMAP7+BAO+$H_0$ dataset. For two-tailed distributions the 68\% CL is given while for one-tailed distributions, the upper 95\% CL is given. In the third and fourth column we list the fiducial values of the various parameters chosen to generate the mock CMB data. `$\Lambda$CDM' corresponds to a cosmology without $\chi$ while in `$\Lambda$CDM$+\chi$', $Y_p$ and $\Neff$ take the values for a 1 MeV Majorana fermion. The last column lists the flat priors we use in $\CosmoMC$.}
\label{tab:forecast}
\end{table}

In the first column of table~\ref{tab:forecast} we list the cosmological parameters that we specify in our forecasts. In addition to $\Omega_{\rm{b}} h^2$, $n_s$, $\Neff$ and $Y_p$ which have previously been defined, we choose values for the cold dark matter density $\Omega_{\rm{DM}} h^2$; the reduced Hubble parameter $h$; the angular scale of the sound horizon at last scattering $\theta_S$;\footnote{In generating the mock data with $\CAMB$, we choose $h$ and $\theta_S$ is derived. In exploring the parameter space with $\CosmoMC$, we sample over $\theta_S$ and $h$ is derived. For completeness we list the values for both parameters.} the optical depth to reionization $\tau$; the amplitude of primordial scalar fluctuations $\ln[10^{10} A_{\rm{S}}]$ (at a pivot scale $k=0.002 \text{ Mpc}^{-1}$) and the fraction of dark matter in the form of neutrinos $f_{\nu}=\Omega_{\nu}/\Omega_{\rm{DM}}$. Finally, we also list the parameter $\sigma_8$, the amplitude of linear matter fluctuations on scales of $8 h^{-1}$ Mpc at $z=0$, which is constrained by galaxy clusters.

In order to find the current best fit values for our parameters, we perform a fit with $\texttt{CosmoMC}$~\cite{Lewis:2002ah} to the SPT+WMAP7+BAO+$H_0$ data. We use v4.1 of the WMAP likelihood code and v1.2 of the SPT likelihood code publicly available from the LAMBDA website~\cite{LAMBDA}. Following~\cite{Keisler:2011aw}, when calculating the SPT likelihood we also sample over three foreground nuisance parameters, in addition to the cosmological parameters listed in table~\ref{tab:forecast}. Here, $Y_p$ is left free, with a top hat prior ranging from 0 to 1. The results of our fit are shown in the second column of table~\ref{tab:forecast}. They are in good accord with similar results found in~\cite{Joudaki:2012fx}.  

To generate mock CMB data, we follow~\cite{Perotto:2006rj,Hamann:2007sb} and use the $\texttt{FUTURCMB}$~\cite{FUTURCMB} add-on package to $\texttt{CosmoMC}$. We use the $\texttt{CAMB}$ software~\cite{Lewis:1999bs} package to obtain angular power spectra for the CMB anisotropies. From this we use the $TT$, $TE$ and $EE$ angular spectra for multiples $2\leq \ell \leq 2500$, and the lensing deflection maps for $dd$ and $Td$. Planck's $B$-mode measurement will likely be noise-dominated so we omit it from our fits.  We assume a sky coverage of $f_{\rm{sky}}=0.65$ and follow~\cite{Hamann:2007sb} in our approach to forecasting by replacing the power spectrum of the mock data by the fiducial data set. Using $\texttt{CosmoMC}$, we generate eight Markov chains in parallel and monitor the convergence with the Gelman-Rubin $R$-statistic \cite{Gelman:1992zz}, requiring that $R-1 \leq 0.015$.

In the third and fourth columns of table~\ref{tab:forecast} we list the fiducial values of the various parameters that we use to generate our mock CMB data. These are chosen to be consistent at or just outside the $1\sigma$ error range of the parameters in the second column. In the third column (`$\Lambda$CDM'), $Y_p$ and $\Neff$ are set to the values predicted from the standard concordance model. In the fourth column (`$\Lambda$CDM$+\chi$'), $Y_p$ and $\Neff$ take the values expected from a 1 MeV Majorana fermion (as determined from figures~\ref{fig:YpDH} and \ref{fig:NeffCMB} respectively). In the fifth column we list the flat priors which we impose on the various parameters when exploring the parameter space with $\CosmoMC$. In this case, we choose a conservative flat prior on $Y_p$. We assume that $Y_p< Y_{\odot}^{\rm{ini}}$, where $Y_{\odot}^{\rm{ini}}$ is the initial solar abundance of helium, for which we take $Y_{\odot}^{\rm{ini}}=0.284$, the upper end of the $1\sigma$ range from~\cite{Serenelli:2010fk}. Our lower bound $Y_p>0.22$ is chosen as it is smaller than all values of $Y_p$ determined from the large data set collated in~\cite{Izotov:2007ed}. Finally, we have chosen parameters to be within $2\sigma$ of the observed value of $\sigma_8$: ref.~\cite{Vikhlinin:2008ym} found $\sigma_8 (\Omega_{\rm{m}}/0.25)^{0.47}=0.813\pm0.013(\rm{stat.})\pm0.024(\rm{syst.})$, where $\Omega_{\rm{m}}$ is the matter density. We have that $\sigma_8 (\Omega_{\rm{m}}/0.25)^{0.47}=0.827$ and $0.855$ for `$\Lambda$CDM' and `$\Lambda$CDM$+\chi$' respectively.

\begin{figure}[t!]
\begin{center}
{
\includegraphics[width=0.985\columnwidth]{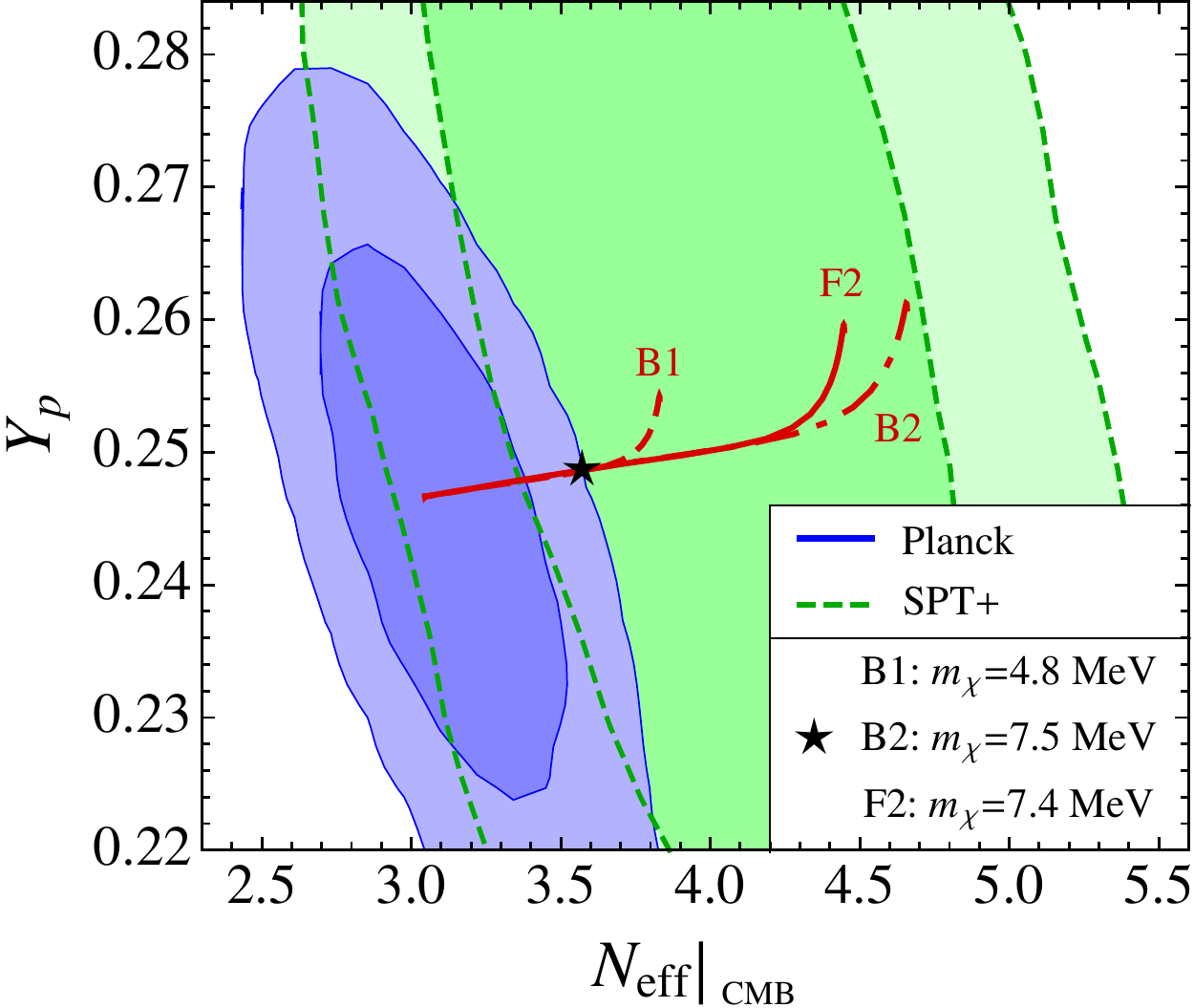}
\includegraphics[width=1.02\columnwidth]{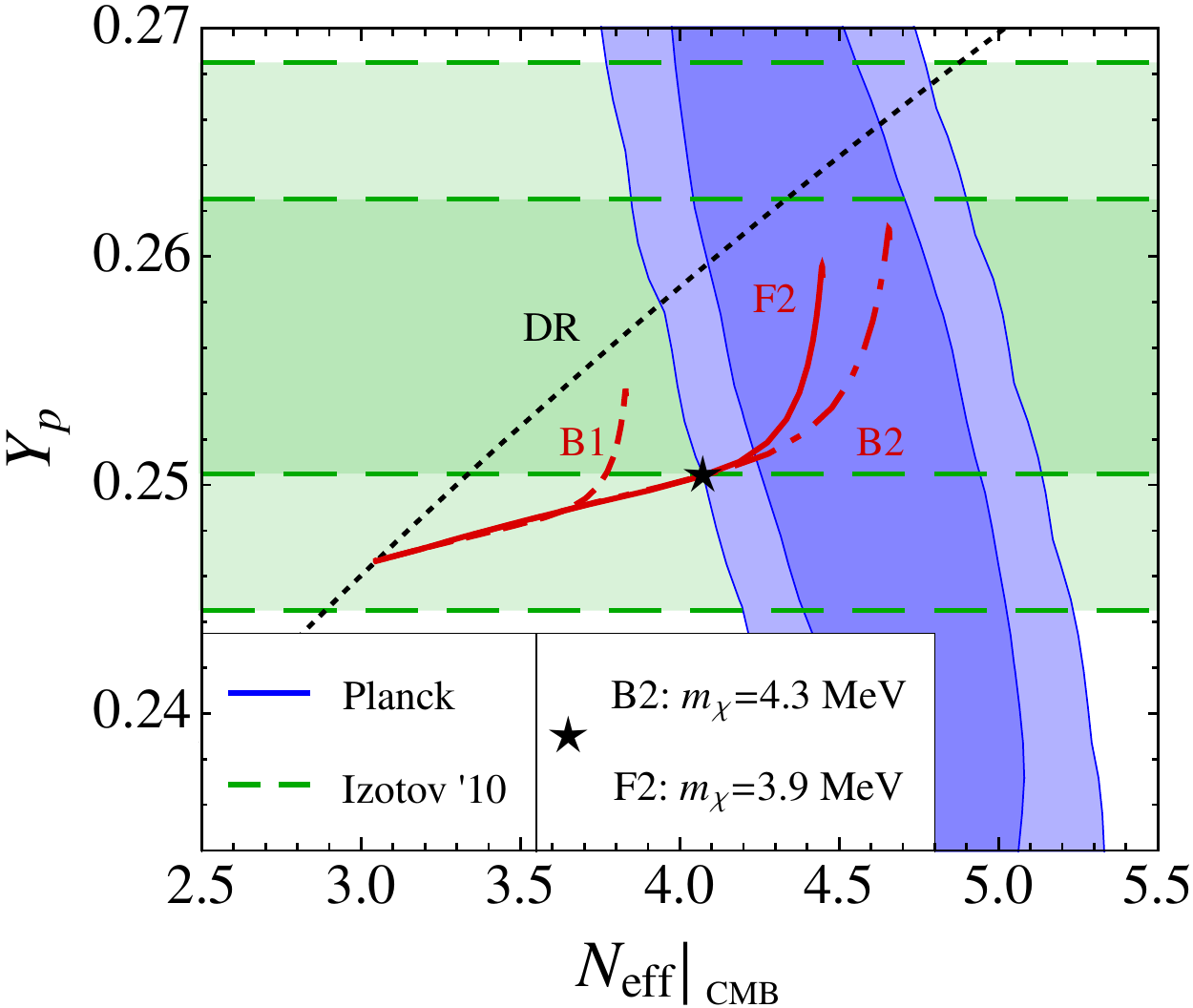}
}
\end{center}
\caption{The upper and lower panels correspond to the fiducial values of `$\Lambda$CDM' and `$\Lambda$CDM$+\chi$' from table~\ref{tab:forecast} respectively. Both panels: In blue are the Planck $1\sigma$ and $2\sigma$ regions. The red lines show the values of $Y_p$ as a function of $\Neff$ for a real scalar (B1), complex scalar (B2) and Majorana fermion (F2). The black star indicates the values of $Y_p$ and $\Neff$ for the stated values of $m_{\chi}$ in the box. Upper panel: In green are the SPT+WMAP7+BAO+$H_0$ $1\sigma$ and $2 \sigma$ regions. Lower panel: In green are the $1\sigma$ and $2 \sigma$ regions for $Y_p$ from Izotov~et.~al.~\cite{Izotov:2010ca} (with statistical and systematic errors combined linearly). The black dotted line marked DR indicates the relation between $Y_p$ and $\Neff$ for dark radiation in which $\NeffBBN=\NeffCMB$.}
\label{fig:Planck}
\end{figure}

In figure~\ref{fig:Planck} we show the results of our forecast in the \mbox{$Y_p$--$\Neff$} plane. The blue contours in the upper and lower panels of figure~\ref{fig:Planck} show the $1\sigma$ and $2\sigma$ regions we find for Planck, after sampling over the mock power spectra created with the `$\Lambda$CDM' and `$\Lambda$CDM$+\chi$' fiducial values from table~\ref{tab:forecast} respectively.\footnote{We do not show the one-dimensional distributions for the individual parameters as we are primarily interested in how well Planck can discriminate between different values of $m_{\chi}$. The one dimensional distributions we find have a similar variance to those found in, for instance \cite{Hamann:2007sb}.}  In both panels, the red dot-dashed, solid and dashed lines show the relation between $Y_p$ and $\Neff$ in the $Y_p$--$\Neff$ plane for a particle $\chi$ when it is a complex scalar (B2), real scalar (B1) or Majorana fermion (F2) respectively. In the lower panel we have adjusted the scale of the vertical axis so as to better resolve these lines. Comparing with figures~\ref{fig:YpDH} and \ref{fig:NeffCMB}, it is clear that the left (right) edge of the lines correspond to the values of $Y_p$ and $\Neff$ for large (small) values of $m_{\chi}$.

In the upper panel of figure~\ref{fig:Planck}, in green we show the $1\sigma$ and $2 \sigma$ regions after fitting to SPT+WMAP7+BAO+$H_0$. Confirming the conclusions from section~\ref{section:CMB}, we see that SPT+WMAP7+BAO+$H_0$ does not place strong constraints on $m_{\chi}$, with all values lying within the $2 \sigma$ region. In comparison, Planck's confidence regions are much smaller and allow much more discrimination between values of $m_{\chi}$. For instance, if Planck makes a measurement of $\Neff$ close to three, as in the upper panel of figure~\ref{fig:Planck}, we see that it is able to exclude at $2\sigma$ values of $m_{\chi}$ below 4.8~MeV for a real scalar, 7.4~MeV for a Majorana fermion and 7.5~MeV for a complex scalar.

In the lower panel of figure~\ref{fig:Planck}, the green regions show the $1\sigma$ and $2\sigma$ regions from the inferred value of $Y_p$ from Izotov~et.~al.~\cite{Izotov:2010ca} with statistical and systematic errors combined linearly. The black dotted line marked DR indicates the usual relation between $Y_p$ and $\Neff$ for dark radiation in which $\NeffBBN=\NeffCMB$. The deviation from this relation for a real scalar, complex scalar and Majorana fermion occurs because $\NeffCMB\geq \NeffYp$. If Planck finds evidence for a larger value of $\Neff$, we see that it will be challenging to distinguish between the case of dark radiation and a generic particle $\chi$ with high statistical significance unless the value of $Y_p$ inferred from astrophysical measurements improves significantly. However, considering only particles in thermal equilibrium with neutrinos, we see that Planck is able to discriminate between a real scalar and a complex scalar or Majorana fermion at more than $2\sigma$ and is able to exclude at $2\sigma$ values of $m_{\chi}$ heavier than 4.3~MeV for a complex scalar and 3.9~MeV for a Majorana fermion.

\section{Conclusions}\label{Section:Conclusion}

Motivated by discrepancies in the determination of $\Neff$ from BBN and the CMB, we have considered the impact of a new light particle which remains in thermal equilibrium with neutrinos until it becomes non-relativistic, which we assume occurs before photon decoupling. Such a particle leads to extra energy density in the early universe, either directly or through its effect on the ratio of the neutrino-to-photon temperature.

To quantify the impact of such particles on BBN, we updated the analysis of \cite{Kolb:1986nf, Serpico:2004nm} through modifications of the $\PArthENoPE$ code to take into account the two effects of light particles on the energy density mentioned above. These particles lead to an increased value of $\Neff$ and bring the $\He$ abundance into better agreement with recent observations (see the upper panel of figure~\ref{fig:YpDH}). We also showed (in the lower panel of figure~\ref{fig:YpDH}) that the $\rm{D}$ abundance is compatible with recent measurements. 

We then considered the effect of such particles at the time of formation of the CMB. At this epoch, the increase in $\Neff$ comes solely from the increase in the neutrino-to-photon temperature ratio, as $\chi$ is non-relativistic. As we showed in figure~\ref{fig:NeffCMB}, this brings the value of $\Neff$ into better agreement with the values reported by CMB experiments. In general, we found that the value of $\Neff$ at the time of CMB formation is larger than the value at BBN. This is demonstrated explicitly in figure~\ref{fig:Neffcomp}. Fixing $Y_p$ to the value inferred from $\Neff$ at photon decoupling, as is often done, is not applicable when setting constraints on this scenario.

As the current experimental constraints on our scenario are still relatively weak, we forecast the sensitivity of the Planck satellite to the effect of these particles on the CMB. We find that Planck is highly sensitive to the effects of such particles. If the value of $\Neff$ derived from Planck agrees with that of standard cosmology with three neutrinos, we showed in the upper panel of figure~\ref{fig:Planck} that it will rule out (for example) the existence of a light Majorana fermion with a mass of less than 7.4 MeV which couples to the neutrinos. The lower panel of figure~\ref{fig:Planck} indicates that it will be difficult to distinguish between our scenario and the standard case of dark radiation (in which $\Neff$ is the same at BBN and photon decoupling) at much more than $1\sigma$ when only data from astrophysical measurements of $\He$ and Planck is considered. On the other hand, considering only particles in thermal equilibrium with neutrinos, we demonstrated in the lower panel of figure~\ref{fig:Planck} that as well as providing significantly improved constraints on $m_{\chi}$, in some regions of parameter space, Planck is even able to discriminate between a real scalar and a complex scalar or Majorana fermion.

\acknowledgments{

We thank Daniel Albornoz V\'asquez, Joanna Dunkley, Jeremy Mardon, Keith Olive and Subir Sarkar for discussions. MJD thanks the theory group at the University of Bonn for hospitality while some of this work was carried out.
}
\appendix

\section{Neutrino-to-photon temperature ratio}\label{Appendix}

The energy density $\rho$ and pressure $p$ of a particle of mass $m$ at temperature $T$ and $g$ internal degrees of freedom are given by (ignoring a possible chemical potential)
\begin{align}
\label{eq:rhop1}
\rho&=\frac{g\, T^4}{2 \pi^2} \int^{\infty}_{y}d\xi \frac{\xi^2 \sqrt{\xi^2-y^2}}{e^{\xi}\pm1}\;,\\
\label{eq:rhop2}
p&=\frac{g \,T^4}{6 \pi^2} \int^{\infty}_{y}d\xi \frac{(\xi^2-y^2)^{3/2}}{e^{\xi}\pm1}\;,
\end{align}
where $\xi=E/T$, $y=m/T$ and here, as usual, $-1$ pertains to bosons and $+1$ to fermions.

In order to calculate the neutrino-to-photon temperature ratio, we use the conservation of entropy per-comoving-volume $S$ for particles in thermal equilibrium with common temperature $T$. That is
\begin{align}
\label{eq:defS}
S=a^3 \,\frac{\rho +p}{T}\equiv a^3\,\frac{2 \pi^2}{45}g_{\star s}T^3=\text{constant}\;.
\end{align}
Here, $a$ is the scale factor and this equation defines $g_{\star s}$, the effective number of relativistic degrees of freedom.

At $T=T_{\rm{D}}$, the neutrinos and $\chi$ (the neutrino plasma) decouple from the electrons and photons (the electromagnetic plasma). From this point onwards, the neutrino and electromagnetic plasmas separately conserve comoving entropy. As a result, the neutrino and photon temperatures ($T_{\nu}$ and $T_{\gamma}$ respectively) evolve according~to
\begin{align}
T_{\nu}&=k_1 \;g_{\star s: \nu}^{-1/3} a^{-1}\;,\\
T_{\gamma}&=k_2\; g_{\star s:\gamma}^{-1/3} a^{-1}\;,
\end{align}
where $k_1$ and $k_2$ are constants and $g_{\star s: \nu}$ and $g_{\star s: \gamma}$ are the effective number of relativistic degrees of freedom in the neutrino and electromagnetic plasmas respectively. Assuming that the neutrinos and $\chi$, and the photons and electrons decouple instantly at $T_{\nu}=T_{\gamma}=T_{\rm{D}}$ (and ignoring that muon and tau neutrinos decouple slightly earlier than the electron neutrino~\cite{Iocco:2008va}), we can determine $k_1/k_2$ and therefore, we find (for $T_{\gamma}<T_{\rm{D}}$)
\begin{equation}
\frac{T_{\nu}}{T_{\gamma}}=\left(\left. \frac{g_{\star s: \nu}}{g_{\star s: \gamma}}\right|_{T_{\rm{D}}} \frac{g_{\star s: \gamma}}{g_{\star s: \nu}} \right)^{1/3}\;,
\label{eq:Tratio1}
\end{equation}
where $\left.\right|_{T_{\rm{D}}}$ indicates that $g_{\star s}$ should be evaluated at $T_{\rm{D}}$.

Using eqs.~\eqref{eq:rhop1}~and~\eqref{eq:rhop2}, the entropy per comoving volume for $N_{\nu}$ neutrinos and $\chi$ is 
\begin{equation}
\label{eq:Snu}
S_{\nu:\chi}=a^3\, \frac{2 \pi^2}{45}T_{\nu}^3 \cdot 2 \cdot \frac{7}{8}\cdot\left[ N_{\nu}+ \frac{g_{\chi}}{2} F(y_{\nu})\right],
\end{equation}
where $y_{\nu}\equiv m_{\chi}/T_{\nu}$ and for convenience, we have defined the function
\begin{equation}
\label{eq:F1}
F(y)=\frac{30}{7 \pi^4}\int^{\infty}_y d\xi \frac{(4 \xi^2-y^2)\sqrt{\xi^2-y^2}}{e^{\xi}\pm1}\;
\end{equation}
with limits, $F(y\rightarrow\infty)= 0$ and \mbox{$F(y\rightarrow0)= 1 \,(8/7)$} for fermions (bosons). Comparing eq.~\eqref{eq:Snu} with eq.~\eqref{eq:defS}, it is straightforward to read off that 
\begin{equation}
g_{\star s:\nu}= 2 \cdot \frac{7}{8}\cdot\left[ N_{\nu}+ \frac{g_{\chi}}{2} F(y_{\nu})\right]\;.
\label{gstarsnu}
\end{equation}
A similar argument shows that
 \begin{equation}
g_{\star s:\gamma}= 2+ 4\cdot\frac{7}{8}\cdot F(z_e)\;,
\label{gstarsgamma}
\end{equation}
where $z_e\equiv m_e/T_{\gamma}$.

\bibliography{ref}
\bibliographystyle{ArXiv}

\end{document}